\newcommand{\eh}[1]{\,\mathrm{#1}}

\newcommand{\gev}{\eh{GeV}}

\newcommand{\tev}{\eh{TeV}}

\newcommand{\volt}{\eh{V}}
\newcommand{\amp}{\eh{A}}

\newcommand{\ohm}{\eh{\Omega}}

\newcommand{\pct}{\eh{\%}}

\newcommand{\pcnt}{$\eh{\%}$}

\renewcommand{\epsilon}{\varepsilon}

\newcommand{\hess}{H.E.S.S.}
\newcommand{\hessI}{H.E.S.S.~I}
\newcommand{\hessIu}{H.E.S.S.~I Upgrade}

\let\seriesfb\bfseries\def\bfseries{\boldmath\seriesfb}
\let\seriesdm\mdseries\def\mdseries{\unboldmath\seriesdm}

\newcommand{\fref}[1]{Fig.~\ref{#1}}

\documentclass{PoS}

\title{A major electronics upgrade for the \hess\ Cherenkov telescopes 1-4}

\ShortTitle{\hessI\ Camera Upgrade}

\author{
\speaker{G.~Giavitto} $^a$,
T.~Ashton~$^b$,
A.~Balzer~$^c$,
D.~Berge~$^c$,
F.~Brun~$^d$,
T.~Chaminade~$^d$,
E.~Delagnes~$^d$,
G.~Fontaine~$^f$,
M.~F\"u{\ss}ling~$^a$,
B.~Giebels~$^f$,
J.F.~Glicenstein~$^d$,
T.~Gr\"aber~$^a$,
J.A.~Hinton~$^{b,g}$,
A.~Jahnke~$^g$,
S.~Klepser~$^a$,
M.~Kossatz~$^a$,
A.~Kretzschmann~$^a$,
V.~Lefranc~$^{a,e}$,
H.~Leich~$^a$,
H.~L\"udecke~$^a$,
P.~Manigot~$^f$,
V.~Marandon~$^g$,
E.~Moulin~$^d$,
M.~de~Naurois~$^f$,
P.~Nayman~$^e$,
M.~Penno~$^a$,
D.~Ross~$^b$,
D.~Salek~$^c$,
M.~Schade~$^a$,
T.~Schwab~$^g$,
R.~Simoni~$^c$,
C.~Stegmann~$^a$,
J.~Thornhill~$^b$,
F.~Toussenel~$^e$ \\
\llap{$^a$}DESY, D-15738 Zeuthen, Germany \\
\llap{$^b$}Department of Physics and Astronomy, The University of Leicester, University Road, Leicester, LE1 7RH, United Kingdom \\
\llap{$^c$}GRAPPA, Anton Pannekoek Institute for Astronomy, University of Amsterdam,  Science Park 904, 1098 XH Amsterdam, The Netherlands \\
\llap{$^d$}DSM/Irfu, CEA Saclay, F-91191 Gif-Sur-Yvette Cedex, France \\
\llap{$^e$}LPNHE, Universit\'e Pierre et Marie Curie Paris 6, Universit\'e Denis Diderot Paris 7, CNRS/IN2P3, 4 Place Jussieu, F-75252, Paris Cedex 5, France \\
\llap{$^f$}Laboratoire Leprince-Ringuet, Ecole Polytechnique, CNRS/IN2P3, F-91128 Palaiseau, France \\
\llap{$^g$}Max-Planck-Institut f\"ur Kernphysik, P.O. Box 103980, D 69029 Heidelberg, Germany \\
Email: \email{gianluca.giavitto@desy.de}, \email{stefan.klepser@desy.de}
}

\abstract{The High Energy Stereoscopic System (\hess) is an array of imaging
atmospheric Cherenkov telescopes (IACTs) located in the Khomas Highland in
Namibia. It consists of four 12-m telescopes (CT1-4), which started
operations in 2003, and a 28-m diameter one (CT5), which was brought online
in 2012. It is the only IACT system featuring telescopes of different
sizes, which provides sensitivity for gamma rays across a very wide energy
range, from $\sim 30\gev$ up to $\sim 100\tev$. Since the camera
electronics of CT1-4 are much older than the one of CT5, an upgrade is
being carried out; first deployment was in 2015, full operation is planned for
2016. The goals of this upgrade are threefold: reducing the dead time of the
cameras, improving the overall performance of the array and reducing the system
failure rate related to aging. Upon completion, the upgrade will assure the
continuous operation of \hess\	at its full sensitivity until and possibly beyond
the advent of CTA.  In the design of the new components, several CTA concepts and
technologies were used and are thus being evaluated in the field: The upgraded read-out electronics is based on the
NECTAR readout chips; the new camera front- and
back-end control subsystems are based on an FPGA and an embedded ARM
computer; the communication between subsystems is based on standard Ethernet
technologies. These hardware solutions offer good performance, robustness
and flexibility. The design of the new cameras is reported here.}

\FullConference{The 34th International Cosmic Ray Conference,\\
		30 July- 6 August, 2015\\
		The Hague, The Netherlands}

\begin{document}

\section{Motivation of the \hessIu}
The \hess\ Cherenkov telescopes 1-4 (aka CT1-4, or "\hessI\ array") were
installed in Namibia, near the Gamsberg mountain, between 2002 and 2004
\cite{hess}. In
2012, a fifth telescope (CT5) was inaugurated in the middle of the previous
\hessI\ array.  In contrast to CT1-4, this new 28-m telescope was not only
equipped with more modern camera electronics, but also operates at a much lower
threshold of $\sim 30\gev$, and therefore delivers a much higher rate of air
shower triggers. Therefore, as MAGIC \cite{magicupgrade} and VERITAS
\cite{veritasupgrade} in the past years, also \hess\ is undergoing a
substantial hardware upgrade to optimise its performance in its final years.
\\
The main reasons to upgrade the old CT1-4 cameras \cite{oldcameras} are (i) to
reduce the downtime of individual cameras, induced by their ageing and
increasing failure rates of cable connectors and other critical parts, and (ii)
to reduce the dead time of the array when operated in coicidence mode with CT5.
The readout time of the original CT1-4 cameras amounts to about $\sim 450
\eh{\mu s}$, which was acceptable for the typical array trigger rates of
\hessI\ ($200-300\eh{Hz}$). However, if these telescopes trigger in coincidence
with CT5, array trigger rates of $1.5\eh{kHz}$ or more are usual, which can
lead to a substantial fraction of events that are not recorded in CT1-4, which
effectively corresponds to a dead time for \textit{stereoscopic} events that can
amount to $30\pct$ or more.
\\
The way to improve on this situation is to replace the current readout scheme
with a fast, Ethernet-based readout and modern front-end boards built around
the NECTAR analog readout chips \cite{nectar}. To reduce the failure rate,
novel schemes for cabling, ventilation, power supply and pneumatics were
developed. The first upgraded camera is deployed in 2015.
\\
These developments offer the opportunity to improve the performance of the
array. For instance, since the new readout can sustain much higher trigger
rates, it would be possible to reduce the trigger thresholds of the CT1-4,
thereby extending their sensitivity to gamma-ray showers of energies below 100
GeV. Other characteristics of the upgraded trigger and readout could also
improve the performance of CT1-4 at higher energies.

\section{Components and aspects}

\begin{figure}
  \centering
  \includegraphics[height=.4\textheight]{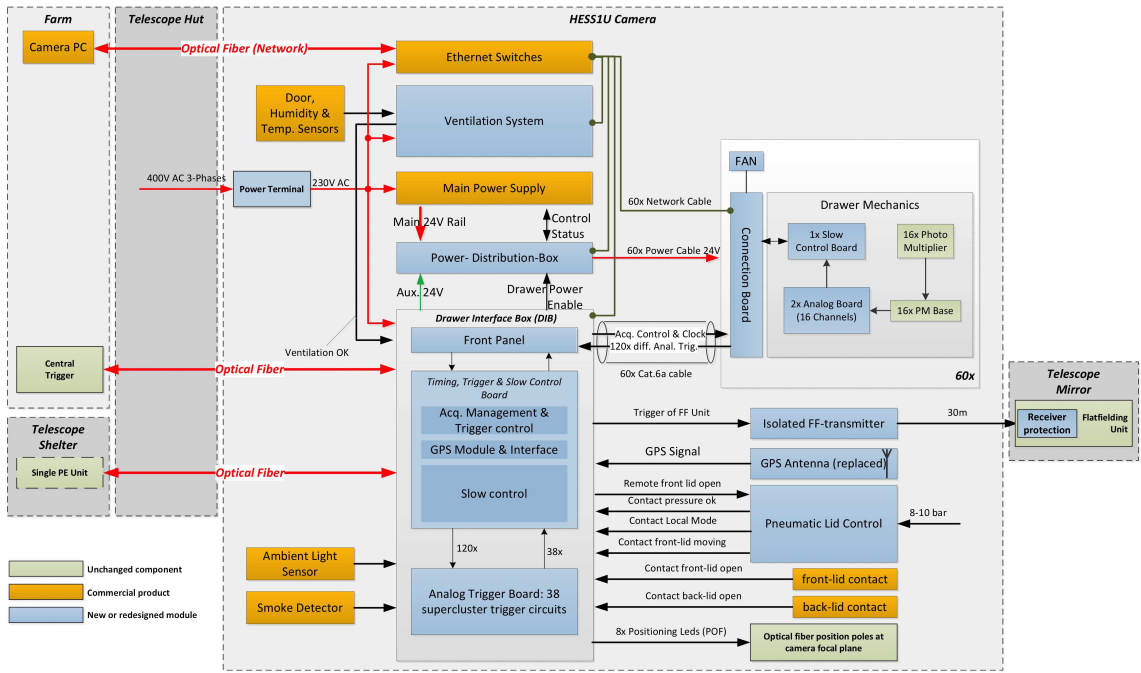}
  \caption{Electronics architecture of the \hessIu. The original \hessI~ systems
      are denoted by green boxes. Where possible, commercial components were
      used (in orange), also in the cabling, where either standard Ethernet or
      M8 cables were deployed. The dashed boxes correspond to physical
      locations: the bigger ``HESS1U Camera`` box corresponds to the camera
      itself, the ``telescope shelter'' is the hut in which the camera is
      parked during the day, the ``telescope hut'' is mounted on the structure
      of the telescope (and co-rotates), the ``farm'' is located inside the
  control building.}
  \label{fig:architecture}
\end{figure}

The upgraded components include electronic and mechanical parts.
Figure~\ref{fig:architecture} summarizes schematically their architecture and
interplay. Basically everything inside the camera has been replaced or
refurbished, except the photomultiplier tubes (PMTs) and their bases, that
generate the high voltage.
\\ 
Following the signal path, from right to left in the figure, the first upgraded
components are the analog front-end boards inside the 60 \emph{drawer} modular
units. Each drawer consists of 16 PMT pixels, whose signals are read out by two
analog boards and is controlled by one slow control board.  The back-end
electronics, which was previously composed of many rack units in two
electronics racks, is now deployed in one rack only. Most of the control,
security and camera trigger logic is implemented on one single rack unit called
\textit{Drawer Interface Box} (DIB).  The drawers are connected to it, to the
data and power networks via a connection board hosting three connectors: two
RJ45 connectors, one for ethernet and one for trigger, readout control and
timing, and one M8 4-pin connector for power (delivered at $24\volt$ DC); The
connection board also hosts the main DC-DC converter of the drawer, in order to
attain a galvanic separation between front- and back-end.
\\
The DIB is composed of 3 boards: a front panel, a main board and an analog
trigger board.  The front connection board is equipped with 60 RJ45 connectors
for the drawers clock, acquisition control and trigger signals, an ethernet
connector, several M8 connectors to sensors and actuators and auxiliary LEMO-00
connectors.  The control and security functions are implemented on the FPGA on
the main board. All peripherical components, like GPS, pneumatics, power
supply, ventilation, are controlled from there, and all the sensors are read
out from there. A security interlock logic on the FPGA firmware evaluates the
sensor inputs: in case of errors or failures it shuts down the power of the
drawers and closes the camera lid.  The main board hosts a separated analog
trigger board, responsible of the camera trigger decision.  
\\
The drawers are powered by the \emph{Power Distribution Box} (PDB), a 64
channel power switch which monitors constantly the current consumption of each
drawer. The main power supply is a commercial 3-module unit which delivers the
$24\volt$ and $\sim80\amp$ needed by the camera. Three modules share the load,
but only two are actually needed, one serves as a hot spare.
\\
The DIB, the power switch and drawers are all equipped with a combination of a
Cyclone IV FPGA and an ARM-based $\mu$Computer Module (Stamp9G45), which allows
for maximum flexibility and ease of control and communication. 

The ventilation concept has been changed completely: a new ventilation system
was implemented on the new back door of the camera, with a single fan,
replacing the many fans previously mounted on the camera sides. Pneumatic
elements were added to be able to open and close the heavier back door, and
the control unit of the entire camera pneumatics was renewed as well.
\\ 
Many sensors were deployed inside and outside the camera for ambient light,
temperature, smoke and humidity. Some of them are included in a new interlock
system to guarantee safe camera operations for both shift crew and hardware.

\subsection{Front-end electronics}

The signal from the PMT base reaches the MCX connector on the analog board via
a $15\eh{cm}$ long coaxial cable, and is there terminated with $R=50\ohm$. The
AC coupled signal is then pre-amplified, split into three branches and further
amplified. Two of these three signal branches are the analog high and low gain,
with total amplification factors 15.1 and 0.68, respectively. They are routed
to the inputs of the NECTAR chip, which have a range of 2V. An adjustable
offset is added to the signals at this point, in order to account for a
possible undershoot and drifts in the baseline due to temperature effects. Its
default value is about $205\eh{mV}$. The third signal branch is amplified by a
factor 45 and routed to a comparator, the output of which goes directly into
the FPGA, where is sampled at $800\eh{MHz}$ and used to generate the trigger.
\\ 
The NECTAR chip is a fast ($1\eh{GS/s}$) sampler and digitizer chip. It
integrates three functions, namely (i) an analogue memory in which the analogue
signal is sampled and stored at high rate, (ii) a 12-bit $21\eh{MHz}$ ADC
digitizing the data stored in the analogue memory and (iii) a serializer, which
sends the digital data downstream to the FPGA.  Each chip integrates two
channels of differential analogue memories with a depth of 1024 cells each.
Operations are simultaneous on the 2 channels. In normal operations only a
small fraction of the cells inside a region of interest (usually 16 cells long)
are read out. The sampling is stopped by an external trigger signal, which must
be synchronous to the analogue signal one whishes to digitize. When reading 16
cells, the nominal dead time of a NECTAR chip is lower than $\sim 2 \eh{\mu
s}$. The way the chip is read out by the FPGA however requires a minimum safe
interval between two events of $\sim 5.5 \eh{\mu s}$, so one can take that
value as a measure of the overall dead time of the camera.
\\
The typical bandwidth of the NECTAR chip analog inputs as reported in its
datasheet \cite{nectardatasheet} is between $300$ and $500\eh{MHz}$. Care was
taken to use analog components matching that value. The overall end-to-end
bandwidth of the readout was measured to be $\sim 330 \eh{MHz}$.
\\
The analog part of the readout was developed to have very low noise. The
pedestal noise, measured as the RMS of the value of a single cell is around 4
ADC counts, or $2\eh{mV}$. At the nominal PMT gain of $2\times10^5$, this
corresponds to 0.32 photoelectrons.
\\
The linearity of the high gain channel is better than 2\pcnt, whereas the low
gain presents correlated deviations from linearity up to 8\pcnt; these errors
can be corrected a posteriori since an absolute calibration of the transfer
functions is performed prior to the installation on the telescope.
\\
The cross talk between two channels within one NECTAR chip and between channels
on different nectar chips is typically less than 0.5\pcnt, and never larger
than 1\pcnt.

\begin{figure}
  \centering
  \includegraphics[height=.35\textheight]{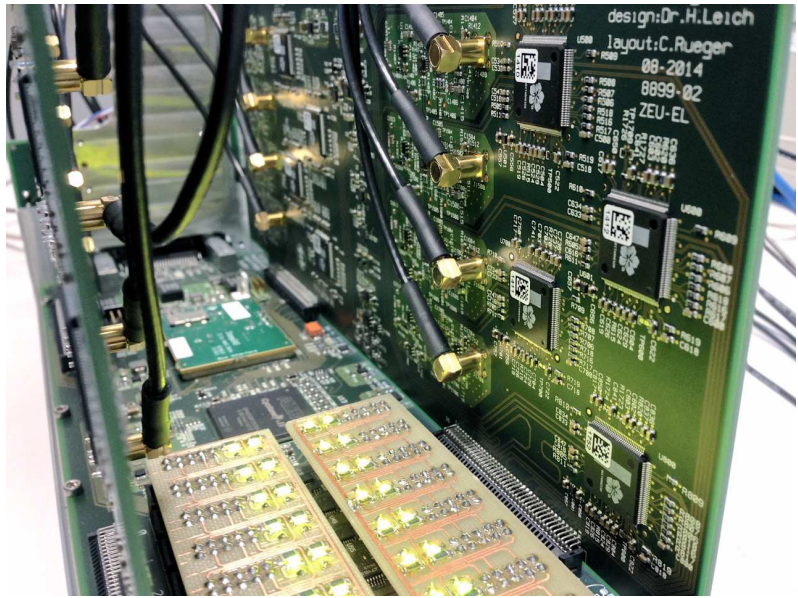}
  \caption{View into a new drawer, with analog front-end boards at the right
and left, and drawer slow control board at the bottom. The analog boards
contain the NECTAR chips (with the flower logo), the slow control board has
test loads mounted at the plugs for the PMT bases (green lights).}
  \label{fig:fe_board_photo}
\end{figure}

\subsection{Trigger chain}

The trigger scheme chosen for the upgraded camera is the same as for the
current cameras \cite{hesstrigger}: An $N$-majority over overlapping sectors of
64 pixels each. As it has been already mentioned, the pixel comparator output
is sampled by the FPGA on the drawer, which in turn generates two analog
signals and sends them to the drawer interface box. These signals carry the
number of pixels above the threshold on its left and right half, respectively;
the number is passed as an analog pulse coded in height in steps of
$33\eh{mV}$. Once
received by the drawer interface box, the signals are isochronously routed to
the 38 overlapping trigger sectors on the analog trigger board, where an analog
sum is built.  The 60 drawers are arranged to occupy the central positions
of a 9x8 matrix, and the sectors overlap in width by one half-drawer and in
height by one full drawer, so the signal from each half drawer is routed to at
most 4 sectors.  The output of the each sector sum goes to a comparator,
where a common threshold signal (corresponding to $N$ pixels fired) is applied.
The 38 comparator outputs are finally combined in an OR logic to give the camera
trigger.
\\ 
The whole trigger path was tested, from the pixel comparator down to the sector
comparator and the camera OR logic. At each position in the trigger chain
dedicated scope measurements were performed to assess the level of noise and
the purity and inefficiency of the trigger. The measurements showed that the
$N$-majority logic implementation has negligible noise and that for every
setting of $N$ exists a threshold setting at which all $N$-multiplicity trigger
are issued and no $N-1$-multiplicity signal can trigger.
\\
The only difference between this design and the original one is the fact that
the trigger comparator output is sampled, albeit at a very high frequency of
800 MHz, whereas in the old design it was not. This allows for greater
flexibility and even for more complex trigger scenarios to be implemented, so
it could work as a testbed for concepts put forward for the next generation of
Cherenkov telescopes (e.g. digital trigger \cite{digitaltrigger}). 
With the present scheme, it is expected that the inefficiency inherent to
sampling these pulses can be mitigated by lowering the pixel thresholds, while
the pixel jitter introduced is negligible (1.25 ns). Extended field studies
are pending.

\subsection{Ventilation}

The new ventilation system, mounted on the back door of the camera, is designed
to steadily inject filtered and (optionally) heated air into the camera. It
causes a steady overpressure in the entire camera body, and leads to a constant
air flow through the drawers and all other small openings of the camera. Like
this, a mild homogeneous cooling is achieved, and dust is constantly prevented
from entering the camera. The option to heat the incoming air can be used to
get rid of humidity after periods of shutdown.
\\
The volume of air exchanged by the ventilation is $>1300\eh{m^3/hr}$
($>360\eh{l/s}$), early tests show that such air flow can keep the drawers at
the required temperature of $\sim 45 \,^{\circ}{\rm C}$, with a spread of $\pm
5 \,^{\circ}{\rm C}$ depending somewhat on the location of the measurement.

\subsection{Network and communication}

Drawers, drawer interface box, power distribution box and ventilation system
are all connected to the main camera server via Ethernet connections.  The
camera network has a star topology, all the aforementioned devices communicate
only with the central camera server and are independent from one another. The
single devices all have 100 Mbit/s copper-based connection to a central switch,
which is then connected to the main camera server by means of a 10 Gbit/s optical
fiber connection. As a novelty, the camera server is now located remotely in the
control building (see \fref{fig:architecture}), easing its maintenance.
\\
The presence of an ARM-based computer module running Linux on all of these
system (with the exception of the ventilation system), makes communication
between single components very easy. In order to reduce the slow control and
data acquisition software development time, several open source software
solutions were adopted. 
\\
The operating system running on the Stamp9G45 is a version of Linux, build
using the Yocto embedded Linux build system \cite{yoctoproject}, with the
kernel module provided by the manufacturer.
\\
Slow control communication is implemented using the Apache Thrift, a
lightweight and performant RPC framework \cite{apachethrift}. 
Data transfer was implemented using the ZeroMQ \cite{zmqsite} smart socket
message-passing library as a transport backbone. Raw event data messages are
serialized via an optimized custom protocol, and monitoring data messages are
serialized using the google protocol buffer library \cite{protobuf}.
\\
These technologies allowed for a robust implementation of network-based slow
control and event acquisition to be developed by a very small team of 2 people in
short timescales. Evaluation tests showed that the Thrift RPC framework can
sustain rates of 10000 requests per second for tens of hours without a single
failure, in a busy, complex network. Bulk data transfer tests using the ZeroMQ
library have reached link saturation on a 1 GB/s connection, and have exceeded
5 GB/s on a 10 GB/s connection. The latter value is still subject to
optimization, and was determined at the time of the test by the CPU speed and
memory available rather than by the protocol implementation.

\subsection{Production tests}

The upgrade requires the production and testing of more than 270 drawers, each
hosting 16 analog channels. The drawers are the most complex item of the
upgrade, therefore an automatic test bench was specifically set up to ease
their testing. An 8-channel pulse generator was designed and build for this
purpose. Each individual channel on one such pulse generator can deliver a
negative pulse with rise and fall times $\sim 1\eh{\rm ns}$, with programmable
attenuation up to -54 dB with 1 dB steps (from $\sim300\eh{\rm mV}$ down to
$\sim0.6\eh{\rm mV}$), programmable delay (up to 64 ns in 0.25 ns steps) and
programmable width (from 2 ns up to 62 ns in 0.25 ns steps). Furthremore the
generator is equipped with a trigger input, a trigger output and a gate, plus a
RJ45 connector to mimick the trigger signal protocol of the upgraded camera.
\\ 
Two of these devices equip the test bench, and permit testing the 16 channels
of a drawer simultaneously. More than 300 single unit tests are performed in a
test run. The tests are grouped in the following categories: basic
functionality, pedestal noise, linearity and cross-talk, and trigger path
tests.  The pedestal noise and trigger path tests are be performed by the ARM
module on drawer itself, since they do not require any external input.
Therefore, it will be possible to test most of the features of a drawer even
after its installation on the camera.

\begin{figure}[t]
  \centering
  \includegraphics[height=.25\textheight]{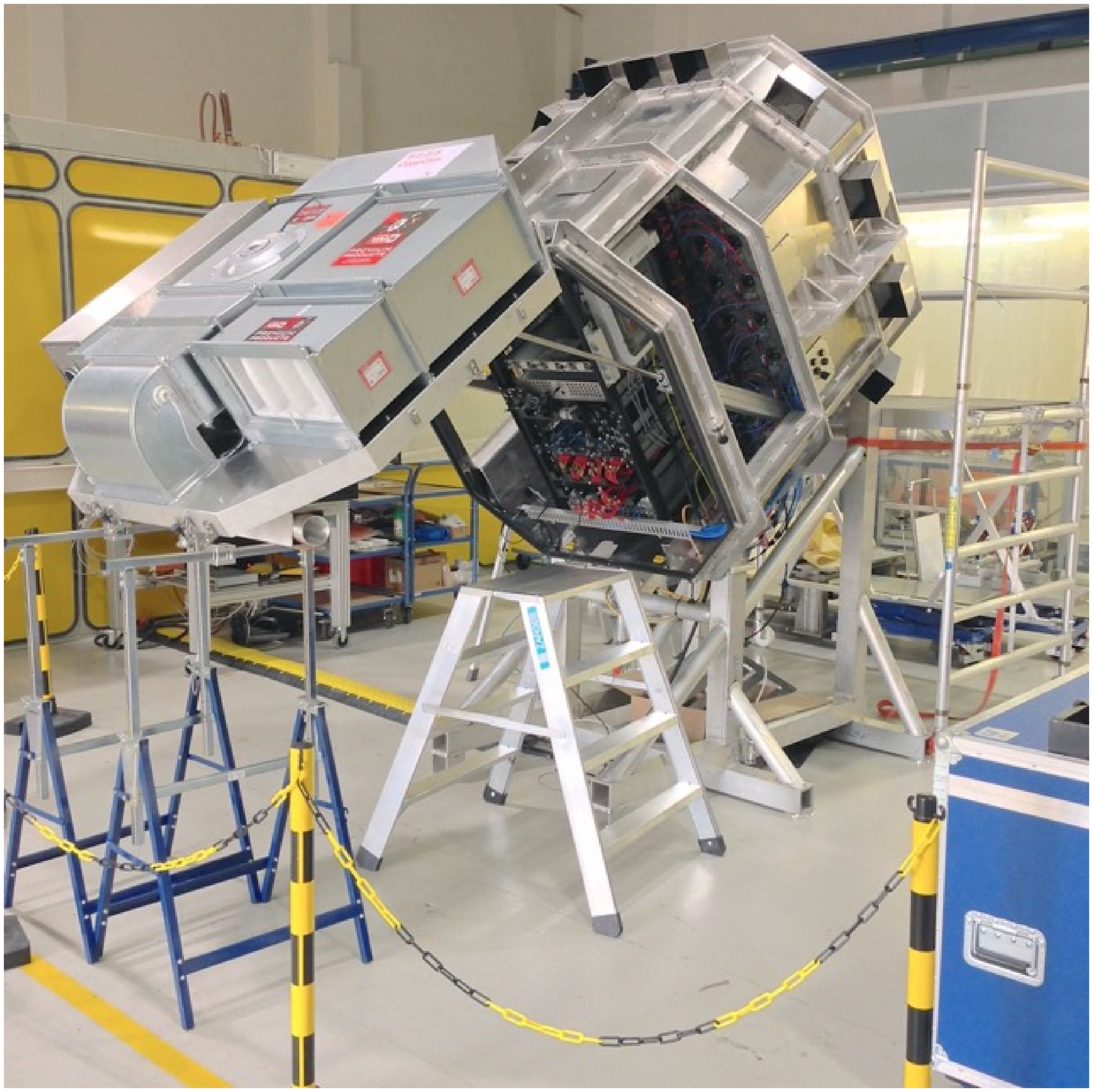}
  \hspace{3mm}
  \includegraphics[height=.28\textheight]{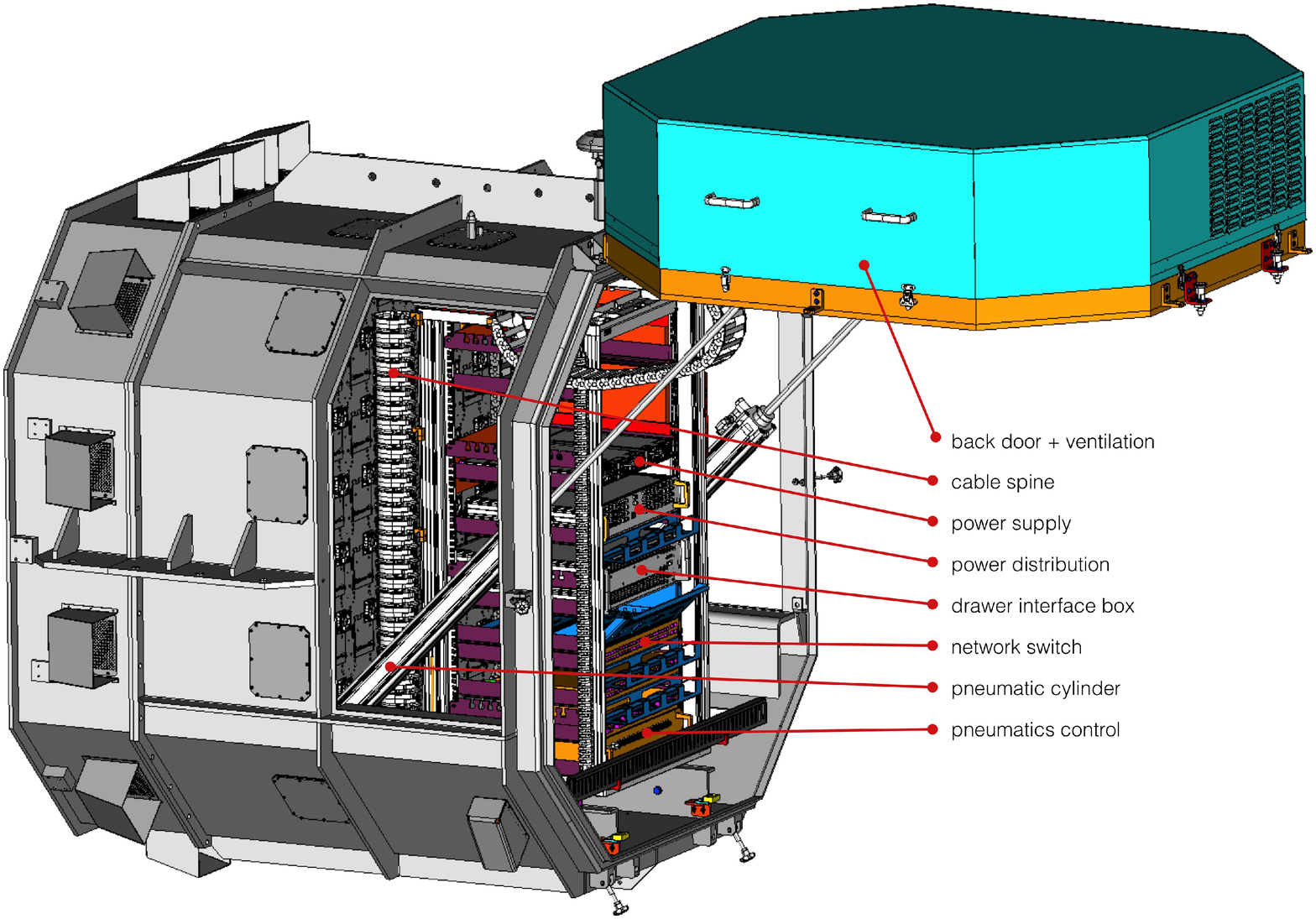}
  \caption{Reproduction of a CT1-4 camera used for test purposes at DESY in
      Zeuthen. Left: Photo with uncovered back door, such that ventilation
      components (filters, fans, heating) can be seen. Inside the camera, the
      rack is mounted with various units installed and blue (trigger) and red
      (data) cables connecting them. The camera is mounted with the same
      inclination as in Namibia, to allow for realistic training of deployment.
      Through the side opening of the camera body, drawer connection boards
      with their fans can be seen.  Right: Labeled sketch of the camera,
      visualising the layout of the rack units} \label{fig:copycam}
\end{figure}

\section{Integration test bench}

In order to test the interplay of the new components as good as possible
before deployment, a replica of a \hessI\ camera was produced and assembled in
Zeuthen (\fref{fig:copycam}). Even without the photomultiplier tubes (PMTs)
available, the trigger chain, camera-internal network, cable mapping,
ventilation, slow control, power supply and mechanical integration could thus
be tested to a good extent.
\\
An additional dark chamber with a 4-drawer ``mini camera'', including PMTs, can
be attached to the test camera to create a realistic dark trigger setup.  A
copy of the \hess\ data acquisition system (DAQ) was attached and the original
state transitions of the system were emulated (see \cite{hessdaq}, Fig.~3).

\section{Outlook and future perspective}

The preliminary plan is to install the first camera on CT1 in 2015, and the
remaining three in one campaign in 2016, after an extensive verification of the
CT1 camera. Besides the improved dead time, the design of the \hessIu\ opens
many possibilities to optimise the trigger logic, front-end readout window and
time information of pulses. The upgraded \hess\ cameras can therefore be
expected to improve the sensitivity of \hess, but also continue to provide a
test bed for various technical innovations that are being discussed in the
framework of CTA R\&D efforts.

\begin{small} \textbf{Acknowledgements.} The support of the Namibian authorities and of the
University of Namibia in facilitating the construction and operation of
H.E.S.S. is gratefully acknowledged, as is the support by the German Ministry
for Education and Research (BMBF), the Max Planck Society, the German Research
Foundation (DFG), the French Ministry for Research, the CNRS-IN2P3 and the
Astroparticle Interdisciplinary Programme of the CNRS, the U.K. Science and
Technology Facilities Council (STFC), the IPNP of the Charles University, the
Czech Science Foundation, the Polish Ministry of Science and Higher Education,
the South African Department of Science and Technology and National Research
Foundation, and by the University of Namibia. We appreciate the excellent work
of the technical support staff in Berlin, Durham, Hamburg, Heidelberg,
Palaiseau, Paris, Saclay, and in Namibia in the construction and operation of
the equipment.
\end{small}

\end{document}